\documentclass{appolb}
\usepackage{epsfig}
\usepackage{hyperref}
\usepackage{url}
\usepackage{calc}
\usepackage{graphicx}

\begin{document}
\title{Bayesian Analysis of the Conditional Correlation Between Stock Index Returns with Multivariate SV Models %
\thanks{Presented at $2$nd Symposium on Socio- and Econophysics, Cracow $21-22$ April $2006$. Research supported by a grant from Cracow University of Economics. The author would like to thank Malgorzata Snarska for help in preparation of the manuscript in Latex format.}%
}
\author{Anna Pajor
\address{Department of Econometrics, Cracow University of Economics}
}
\maketitle
\begin{abstract}
In the paper we compare the modelling ability of discrete-time
multivariate Stochastic Volatility models to describe the
conditional correlations between stock index returns. We consider
four trivariate SV models, which differ in the
 structure of the conditional covariance matrix. Specifications with zero, constant and time-varying
 conditional correlations are taken into account. As an example we study trivariate volatility models for the daily log returns on the WIG,
 S\&P $500$, and FTSE $100$ indexes. In order to formally compare the relative explanatory power of SV specifications
 we use the Bayesian principles of comparing statistic models. Our results are based on the Bayes factors and implemented
 through Markov Chain Monte Carlo techniques. The results indicate that the most adequate specifications are those that
 allow for time-varying conditional correlations and that have as many latent processes as there are conditional
  variances and covariances. The empirical results clearly show that the data strongly reject the assumption of constant
  conditional correlations.
\end{abstract}
\section{Introduction}
There are a lot of theoretical and empirical reasons to study multivariate volatility models. Analysis of financial market
 volatility and correlations among markets play a crucial role in financial decision making (e.g. hedging strategies,
 portfolio allocations, Value-at-Risk calculations). The correlations among markets are very important in the global
 portfolio diversification.\\The main aim of the paper is to compare the modelling ability of discrete-time Multivariate
 Stochastic Volatility (MSV) models to describe the conditional correlations and volatilities of stock index returns.
 The MSV models offer powerful alternatives to multivariate GARCH models in accounting
 for properties of the conditional variances and correlations. Superior performance of bivariate SV models over GARCH
 models (in term of the Bayes factor) are documented in \cite{Osiewalski}. But the MSV models are not as
 often used in empirical applications as the GARCH models. The main reason is that the SV models are more difficult to
 estimate. In this paper we consider four multivariate Stochastic Volatility models, including the specification with zero,
 constant and time-varying conditional correlations. These MSV specifications are used to model volatilities and
 conditional correlations between stock index returns. We study trivariate volatility models for the daily log returns
 on the WIG index, the Standard \& Poor's $500$ index, and the FTSE $100$ index for the period January $4$, $1999$ to December $30$, $2005$.
 In the next section the Bayesian statistical methodology is briefly presented . In section $3$ the model
 framework is introduced. Section $4$ is devoted to the description of trivariate SV specifications. In section $5$ we
 present and discuss the empirical results.
\section{Bayesian statistical methodology}
Let $\mathbf{y}$ be the observation matrix and $\theta_i$ be the
vector of unknown parameters and $\omega_i$ the latent variable
vector in model $M_i$ $(i = 1, 2,\ldots, n)$. The $i$ - the
Bayesian model is characterized by the joint probability density
function, which can be written as the product of three densities:
\[
 p(\mathbf{y}, \omega_i,\theta_i|
y_{(0)}, M_i) = p(\mathbf{y}| \omega_i,\theta_i ,y_{(0)}, M_i)
p(\omega_i
 | \theta_i, M_i) p( \theta_i| M_i), \; i = 1, 2, \ldots, n,
\]where $y_{(0)}$ denotes initial conditions, $ p(\mathbf{y}| \omega_i,\theta_i ,y_{(0)}, M_i)$ is
the conditional density of $\mathbf{y}$ when  $\omega_i \in
\Omega_i, \theta_i \in \Theta_i$ are given, $ p(\omega_i
 | \theta_i, M_i) $ is the density of
the latent variables conditioned on $\theta_i$, $p( \theta_i| M_i)$
is the prior density function under $M_i$. The joint probability
density function can be expressed as the product of the marginal
data density of the observation matrix (given the initial conditions
$y_{(0)}$) in model $M_i$: $p(\mathbf{y}| y_{(0)}, M_i)$, and the
posterior density function of the parameter vector  $\theta_i$ and
the latent variable vector $\omega_i$ in $M_i$: $p(\omega_i,\theta_i
| \mathbf{y}, y_{(0)}, M_i)$, i.e.
\[p(\mathbf{y}, \omega_i, \theta_i| y_{(0)}, M_i) = p(\omega_i,\theta_i
| \mathbf{y}, y_{(0)}, M_i) p(\mathbf{y}| y_{(0)}, M_i),\] where
\[p(\mathbf{y}| y_{(0)},
M_i)=\int_{\Omega_i\times\Theta_i}p(\mathbf{y}|\omega_i, \theta_i,
y_{(0)}, M_i)p(\omega_i, \theta_i|M_i) d\omega_i d\theta_i.\] The
statistical inference is based on the posterior distributions,
while the marginal densities $p(\mathbf{y}| y_{(0)}, M_i)$ $(i =
1, 2, \ldots, n)$ are the crucial components in model
comparison.\\ Assume that $M_1, \ldots, M_n$ are mutually
exclusive (non-nested) and jointly exhaustive models. From Bayes's
theorem, it is easy to show that the posterior probability of
$M_i$ is given by:
\[p(M_i|\mathbf{y}, y_{(0)})=\frac{p(M_i)p(\mathbf{y}|y_{(0)},
M_i)}{\sum_{i=1}^n p(M_i)p(\mathbf{y}|y_{(0)}, M_i)},\] where
$p(M_i)$ denotes the prior probability of $M_i$. For the sake of
pairwise comparison, we use the posterior odds ratio, which for
any two models $M_i$ and $M_j$ is equal to the prior odds ratio
times the ratio of the marginal data densities:
\[\frac{p(M_i|\mathbf{y}, y_{(0)})}{p(M_j|\mathbf{y}, y_{(0)})}=\frac{p(M_i)}{p(M_j)}\cdot\frac{p(\mathbf{y}|y_{(0)},
M_i)}{p(\mathbf{y}|y_{(0)}, M_j)}.\]The ratio of the marginal data
densities is called the Bayes factor:
\[B_{ij}=\frac{p(\mathbf{y}|y_{(0)},
M_i)}{p(\mathbf{y}|y_{(0)}, M_j)}.\] Thus, assuming equal prior
model probabilities (i.e. $p(M_i) = p(M_j)$), the Bayes factor is
equal to the posterior odds ratio. We see that the values of the
marginal data densities for each model are the main quantities for
Bayesian model comparison. The marginal data density in model $M_i$
can be written as:
\[p(\mathbf{y}| y_{(0)},
M_i)=\left(\int_{\Omega_i\times\Theta_i}\left[p(\mathbf{y}|\omega_i,
\theta_i, y_{(0)}, M_i)\right]^{-1}p(\omega_i,
\theta_i|\mathbf{y}, y_{(0)}, M_i) d\omega_i
d\theta_i\right)^{-1}\] Of course, in the case of SV models this
integral can not be evaluated analytically and thus must be
computed by numerical methods. We use the method proposed by
\cite{Newton}, which approximates the marginal data density by the
harmonic mean of the values $p(\mathbf{y}|\omega_i, \theta_i,
y_{(0)}, M_i)$, calculated for the observed matrix $\mathbf{y}$
and for the vector $(\omega_ i^{(q)} , \theta_i^{(q)})'$ drawn
from the posterior distribution. That is:
\[\hat{p}(\mathbf{y}|y_{(0)},
M_i)=\left(\frac{1}{m}\sum_{q=1}^m\frac{1}{p(\mathbf{y}|\omega_i^{(q)},
\theta_i^{(q)}, y_{(0)}, M_i)}\right)^{-1}.\] The estimator
$\hat{p}(\mathbf{y}|y_{(0)}, M_i)$ is very easy to calculate and
gives results that are precise enough for our model comparison.
\section{Model framework}
Let $x_{jt}$ denote the price of asset $j$ (or index quotations as
in our application) at time $t$ for $j = 1, 2, 3$ and $t = 1, 2,
\ldots , T$. The vector of growth rates $y_t =(y_{1,t}, y_{2,t},
y_{3,t})'$, each defined by the formula $y_{j,t} = 100 \ln
\left(x_{t,j}/x_{j,t-1}\right)$, is modelled using the VAR(1)
framework:
\[ y_t -\delta = R(y_{t-1} - \delta) + \xi_t , \;\;\; t = 1, 2,\ldots ,T,\]
where $\{\xi_t\}$ is a trivariate SV process, $T$ denotes the number
of the observations used in estimation. More specifically:
\[\left[
    \begin{array}{c}
      y_{1,t} \\
      y_{2,t} \\
      y_{3,t} \\
    \end{array}
  \right] - \left[
    \begin{array}{c}
      \delta_1 \\
      \delta_2  \\
      \delta_3  \\
    \end{array}
  \right]= \left[
              \begin{array}{ccc}
                r_{11} & r_{12} & r_{13} \\
                r_{21} & r_{22} & r_{23} \\
                r_{31} & r_{32} & r_{33} \\
              \end{array}
            \right]\left(\left[
    \begin{array}{c}
      y_{1,t-1} \\
      y_{2,t-1} \\
      y_{3,t-1} \\
    \end{array}
  \right] - \left[
    \begin{array}{c}
      \delta_1 \\
      \delta_2  \\
      \delta_3  \\
    \end{array}
  \right]\right)+ \left[\begin{array}{c}
             \xi_{1,t} \\
             \xi_{2,t} \\
             \xi_{3,t} \\
           \end{array}\right].\]
We assume that, conditionally on the latent variable vector
$\Omega_{t(i)}$ and on the parameter vector $\theta_i, \xi_t$
follows a trivariate Gaussian distribution with mean vector $0_{[3
\times 1]}$ and covariance matrix $\Sigma_t$, i.e.
\[ \xi_t|\Omega_{t(i)}, \theta_i \sim N(0_{[3 \times 1]},\Sigma_t),\;\; t = 1, 2, \dots , T.\]
Competing trivariate SV models are defined by imposing different
structures on $\Sigma_t$.\\For all elements of $\delta$  and $R$
we assume the multivariate standardized Normal prior $N(0,
I_{15})$, truncated by the restriction that all eigenvalues of $R$
lie inside the unit circle. We assume that the matrix $[\delta ,
R]$ and the remaining (model-specific) parameters are prior
independent.
\section{Trivariate VAR(1) - SV models}
\subsection{Stochastic Discount Factor Model (SDF)}
The first specification considered here is the stochastic discount
factor model (SDF) proposed, but not applied, by \cite{Jacquier}.
The SDF process is defined as follows:
\[\begin{array}{rcl} \xi_t = \varepsilon_t\sqrt{h_t}, &&\varepsilon_t \sim iiN(0_{[3 \times 1]},\Sigma), \\
\ln h_t =\phi\ln h_{t-1} +\sigma_h\eta_t, && \eta_t \sim iiN(0,1),\\
\varepsilon_{j,t} \perp \eta_s, && t,s \in \mathbf{Z},\;j = 1,2,3,
\end{array}\]
 where $\mathbf{Z} = \{\dots, -2, -1, 0, 1, 2, \ldots\}$, $\perp$  denotes independence, and the symbol $\eta_t \sim
 iiN(0,1)$
denotes a series of independently and normally distributed random
variables with mean vector $0_{[3\times 1]}$ and covariance matrix
$\Sigma$. In this case, we have
\[\xi_t|\Omega_{t(1)}, \Sigma \sim  (0_{[3 \times 1]},h_t\Sigma),\]
where $\Omega_{t(1)}=h_t$. The conditional covariance matrix of
$\xi_t$ is time varying and stochastic, but all its elements have
the same dynamics governed by $h_t$. Consequently, the conditional
correlation coefficients are constant over time. Our model
specification is completed by assuming the following prior
structure:
\[p(\phi, \sigma_h^2, \ln h_0 , \Sigma) = p(\phi)p(\sigma_h^2)p(\ln h_0)p(\Sigma),\]
where we use proper prior densities of the following
distributions: \\ $ \phi \sim N(0, 100)I_{(-1,1)}(\phi),\;
\;\sigma_h^2 \sim IG(1, 0.005),\;\; \ln h_0 \sim N(0, 100),\; \\
\Sigma \sim IW(3I, 3, 3)$. \\ The symbol $N(a, b)$ denotes the
normal distribution with mean $a$ and variance $b$, $I_{(-1,
1)}(.)$ is the indicator function of the interval $(-1, 1)$.
$IG(\nu_0, s_0)$ denotes the inverse Gamma distribution with mean
$s_0/(\nu_0-1)$ and variance $s_0^2/[(\nu_0 -1)^2(\nu_0 -2)]$. The
symbol $IW(B, d, 3)$ denotes the three-dimensional inverse Wishart
distribution with $d$ degrees of freedom and parameter matrix $B$.
The initial condition for $\ln h_t$ (i.e. $\ln h_0$) is treated as
an additional parameter and estimated jointly with other
parameters.
\subsection{Basic Stochastic Volatility Model (BSV)}
Next, we consider the basic stochastic volatility process (BSV),
where $\xi_t |\Omega_{t(2)} \sim N(0_{[3 \times 1]},\Sigma_t),$
and $\Sigma_t = Diag(h_{1,t}, h_{2,t}, h_{3,t})$ (similar to the
idea of \cite{Harvey}). The conditional variance equations are: \[
\ln h_{j,t}  - \gamma _{jj} =\phi_{jj}(\ln h_{j,t-1}  - \gamma
_{jj}) + \sigma_{jj}\eta_{j,t}, \] for $j = 1, 2, 3$, where
$\eta_t \sim iiN(0_{[3 \times 1]},I_3)$, $\eta_t =(\eta_{1,t},
\eta_{2,t},\eta_{3,t})'$, $\Omega_{t(2)}=(h_{1,t},
h_{2,t},h_{3,t})'$. For the parameters we use the same
specification of prior distribution as in the univariate SV model
(see \cite{Pajor:2003}), i.e. $ (\gamma_{jj}, \phi_{jj})' \sim
N(0, 100I)I_{(-1,1)}(\phi_{jj}),\; \;\sigma_h^2 \sim IG(1,
0.005),\;\; \ln h_{j,0} \sim N(0, 100),\;\; j= 1,2,3.$
\subsection{JSV Model} Both previous specifications (SDF and BSV) are
very restrictive. Now, we propose a SV process based on the spectral
decomposition of the matrix $\Sigma_t$. That is
\[ \Sigma_t = P \Lambda_t P^{-1},\]
where $\Lambda_t$  is the diagonal matrix consisting of all
eigenvalues of $\Sigma_t$, and  $P$ is the matrix consisting of
the eigenvectors of $\Sigma_t$. For series $\{\ln \lambda_{j,t}\}$
$(j = 1, 2, 3)$, similarly as in the univariate SV process, we
assume standard univariate autoregressive processes of order one,
namely
\[ \ln\lambda_{j,t} -\gamma_{jj} = \phi_{jj}(\ln \lambda_{j,t-1} - \gamma_{jj}) +\sigma_{jj}\eta_{j,t}, \]
for $j = 1, 2, 3$ , where $\eta_t \sim iiN(0_{[3 \times 1]},I_3)$,
$\eta_t =(\eta_{1,t}, \eta_{2,t},\eta_{3,t})'$, and
$\Omega_{t(3)}=(\lambda_{1,t}, \lambda_{2,t},\lambda_{3,t})'$.
This reparametrization of $\Sigma_t$ does not require any
parameter constraints to ensure positive definiteness of
$\Sigma_t$. If $|\phi_{jj}| < 1 \;\; (j=1, 2,3)$ then $\{\ln
\lambda_{1,t}\}$, $\{\ln \lambda_{21,t}\}$ and $\{\ln
\lambda_{3,t}\}$ are stationary and the JSV process is a white
noise. In addition, $P$ is an orthogonal matrix, i.e. $P'P=I_2$,
thus $P$ is parametrized by three parameters (Euler angles)
$\kappa_j \in (- \pi ,\pi )$, $j \in \{ 1, 2, 3\}$:
\[P(\kappa_1, \kappa_2, \kappa_3) =
P_1(\kappa_1)P_2(\kappa_2)P_3(\kappa_3),\]
where for $l = 1, 3$
\[ P_l (\kappa_l) = \left[%
\begin{array}{ccc}
\cos\kappa_l & -\sin\kappa_l & 0 \\
\sin\kappa_l & \cos\kappa_l & 0 \\
0 &0 & 1
\end{array}%
\right], \quad P_2(\kappa_2) = \left[%
\begin{array}{ccc}
1 &0 & 0\\
0 &\cos\kappa_2 & -\sin\kappa_2 \\
0 & \sin\kappa_2 & \cos\kappa_2
\end{array}%
\right]. \] In this case the conditional correlation coefficients
are time-varying and stochastic if $\kappa_j\neq 0$ for some $j
\in \{ 1, 2, 3\}$. For the model-specific parameters we take the
following prior distributions: $ (\gamma_{jj}, \phi_{jj})' \sim
N(0, 100I)I_{(-1,1)}(\phi_{jj}),\; \\ \sigma_{jj}^2 \sim IG(1,
0.005),\;\; \ln \lambda_{j,0} \sim N(0, 100)$, $\kappa_j \sim
U(-\pi ,\pi)$ (i.e. uniform over $(-\pi ,\pi)$), $j = 1, 2, 3$.
The BSV model can be obtained by imposing the parameter
restrictions $\kappa_1 = \kappa_2 = \kappa_3 = 0$ in the $P$
definition of the JSV model (but we formally exclude this value).
\subsection{TSV Model} The next specification (proposed by \cite{Tsay},
thus called TSV) uses six separate latent processes (the number of
the latent processes is now equal to the number of distinct elements
of the conditional covariance matrix). Following the definition in
\cite{Tsay}, we propose to use the Cholesky decomposition:
 \[\Sigma_t = L_t G_t L_t',\]
where $L_t$ is a lower triangular matrix with unitary diagonal
elements, $G_t$ is a diagonal matrix with positive diagonal
elements:
\[ L_t = \left[%
\begin{array}{ccc}
  1 & 0 & 0  \\
  q_{21,t} & 1 & 0  \\
  q_{31,t} & q_{32,t} & 1  \\
\end{array}%
\right] , \quad G_t = \left[%
\begin{array}{ccc}
  q_{11,t} & 0 & 0  \\
  0 & q_{22,t} & 0  \\
  0 & 0 & q_{33,t}  \\
\end{array}%
\right], \] that is
\[\Sigma_t = \left[%
\begin{array}{ccc}
  q_{11,t} & q_{11,t}q_{21,t} & q_{11,t}q_{31,t}  \\
  q_{21,t}q_{11,t} & q_{11,t}q_{21,t}^2 + q_{22,t} & q_{11,t}q_{21,t}q_{31,t} + q_{22,t}q_{32,t} \\
  q_{31,t}q_{11,t} & q_{11,t}q_{21,t}q_{31,t} + q_{22,t}q_{32,t}& q_{11,t}q_{31,t}^2 + q_{22,t}q_{32,t}^2 + q_{33,t} \\
\end{array}%
\right].\] Series $\{q_{ij,t}\}$, and $\{\ln q_{jj,t}\}$ $(i, j =
1, 2, 3,\quad i > j)$, analogous to the univariate SV, are
standard univariate autoregressive processes of order one, namely
\[ \ln q_{jj,t} -\gamma_{jj} =
\phi_{jj}(\ln q_{jj,t-1} - \gamma_{jj})
+\sigma_{jj}\eta_{jj,t},\;\;\: j = 1, 2, 3,\] \[q_{ij,t}
-\gamma_{ij} = \phi_{ij}(q_{ij,t-1} - \gamma_{ij})
+\sigma_{ij}\eta_{ij,t},\quad j, i  \in \{1, 2, 3\},\quad i > j,\]
\[\eta_t = (\eta_{11,t}, \eta_{22,t}, \eta _{33,t},\eta_ {21,t},
\eta_{31,t}, \eta_{32,t})' \sim iiN_6(0_{[6\times 1]}, I_6),\quad
t \in \mathbf{Z},\] \[ \Omega_{t(4)} = (q_{11,t}, q_{22,t},
q_{33,t}, q_{21,t}, q_{31,t}, q_{32,t})'.\] Note that positive
definiteness of $\Sigma_t$ is achieved by modelling $\ln q_{jj,t}$
instead of $q_{jj,t}$ . It is easy to show that if the absolute
values of $\phi_{ij}$ are less than one the TSV process is a white
noise (see \cite{Pajor:2005a}). We see that the TSV model is able
to model both the time-varying conditional correlation
coefficients and variances of returns. A major drawback of this
process is that the conditional variances and covariances are not
modelled in a symmetric way, thus the explanatory power of model
may depend on the ordering of financial instruments.\\We assume
the following prior distributions: \\$ (\gamma_ {ij},\phi_{ij})'
\sim N(0, 100I) I_{(-1,1)}(\phi_{ij})$, $\sigma_{ij}^2 \sim IG(1,
0.005)$, $\ln q_{ii,0} \sim N(0, 100)$ for $i, j \in \{1, 2, 3\}$
and $i \geq j$ ; $q_{ij,0} \sim N(0, 100)$ for $i, j \in \{1, 2,
3\}$, $i > j$. The prior distributions used are relatively
noninformative. Note that the BSV model can be obtained as a
limiting case, corresponding to $ \gamma_{ij} = \phi_{ij} = 0,$
$\sigma_{ij}^2 \to 0$ for $i, j \in \{1, 2, 3\}$, $i
> j$.

\section{Empirical results}
We consider daily stock index returns for three national markets:
Poland (WIG), the United States (S\&P $500$), and the United
Kingdom (FTSE $100$), from January $4, 1999$ to December $30,
2005$. We consider only index closing quotations in trading days
for all considered national markets, thus our sample consists of
$1701$ daily observations \footnote {The data were downloaded from
the websites \url{(http://finance.yahoo.com)} and
\url{http://www.parkiet.com/dane/dane_atxt.jsp}, where complete
descriptions of the indices can be found.}. The first observation
is used to construct initial conditions. Thus T, the length of the
modelled vector time series, is equal to $1700$.
\begin{table}
  \centering
  \caption{Logs of Bayes factors in favour of $TSV_{FSW}$ model}\label{Table:1}
\begin{tabular}{|c|c|c|c|c|}
  \hline
  &&&&\\
  Model & Number of latent   & Number of & $\log_{10} (B_{4,1,i})$ & Rank \\
   & processes  & parameters&  &
   \\ &&&& \\ \hline
   $M_{4,1}$ $ (TSV_{FSW})$& 6 &  12+24 & 0.00  &   1 \\ \hline
   $M_{4,2}$ $ (TSV_{FWS})$& 6 &  12+24 & 7.82  &   2 \\ \hline
   $M_{4,3}$ $ (TSV_{SWF})$& 6 &  12+24 & 15.55 &   3 \\ \hline
   $M_{4,4}$ $ (TSV_{SFW})$& 6 &  12+24 & 15.86 &   4 \\ \hline
   $M_{4,5}$ $ (TSV_{WFS})$& 6 &  12+24 & 17.05 &   5 \\ \hline
   $M_{4,6}$ $ (TSV_{WSF})$& 6 &  12+24 & 22.96 &   6 \\ \hline
  $M_3$ (JSV) & 3 & 12+15 & 63.68 & 7 \\ \hline
  $M_1$ (SDF) & 1 & 12+9 & 87.39 & 8 \\ \hline
  $M_2$ (BSV) & 3 & 12+12 & 181.18 & 9 \\
  \hline
\end{tabular}
\end{table}
In Table 1 we present the decimal logarithms of the Bayes factors
in favor of $TSV_{FSW}$ model. Our posterior results are obtained
using MCMC methods: Metropolis-Hastings within the Gibbs sampler
(see \cite{Pajor:2006}, \cite{Hagan} and \cite{Gamerman}). The
results presented in this paper are based on $500,000$ states of
the Markov chain, generated after $100,000$ burnt-in states. The
Bayes factors are calculated using the Newton and Raftery's method
\cite{Newton}. Because in the TSV specification the conditional
variances are not modelled in a symmetric way, we consider six
cases: $TSV_{FSW}$, $TSV_{FWS}$, $TSV_{SWF}$, $TSV_{SFW}$,
$TSV_{WFS}$, and $TSV_{WSF}$. These models differ in ordering of
elements in $y_t$. For example in the $TSV_{FSW}$ model $y_{1,t}$
denotes the daily growth rate of the FTSE $100$ index at day $t$,
and $y_{2,t}$ and $y_{3,t}$ are respectively the daily growth
rates of the S\&P $500$ and the WIG indexes at day $t$.\\Our
findings show clear superiority of the TSV specifications (which
describe the six distinct elements of the conditional covariance
matrix by six separate latent processes) over all SV models
considered here. The $TSV_{FSW}$ model receives almost all
posterior probability mass (assuming equal prior model
probabilities), being about $7.82$ orders of magnitude more
probable a posterior than the $TSV_{FWS}$ model and $63.68$ orders
of magnitude better than the JSV model. Furthermore, the
$TSV_{WSF}$ model fits the data about $23$ orders of magnitude
worse than the best TSV model. It is mainly attributed to the fact
that the growth rates of the FTSE index are less volatile than the
S\&P and WIG indexes. When we compare the unconditional variance
of $\xi_ {j,t}$ $(Var(\xi_{j,t}) = \exp\left( \gamma_{jj} + 0.5
\sigma_{jj}^2/(1- \phi_{jj}^2)\right),\;\; j = 1, 2, 3)$ obtained
in the BSV model, we observe a value of $1.448$ for the WIG index,
$0.955$ for the S\&P $500$ index and $0.943$ for the FTSE index.
It is in accordance with the ordering of returns in the best TSV
model. Thus, the explanatory power of the SV model depends not
only on the number of latent processes, but also on the ordering
of financial instruments in case of the TSV specifications. The
results indicate that the return rates of the WIG, S\&P and FTSE
indexes reject the constant or zero conditional correlation
hypothesis, represented by the SDF and BSV model.\\The main
characteristics of the posterior distributions of the conditional
correlation coefficients are presented in Figure 1, where the
upper line represents the posterior mean plus standard deviation
the lower one - the posterior mean minus standard deviation. The
conditional correlation coefficients produced by our VAR(1)-SV
models with at least three latent processes vary markedly over
time. Surprisingly, the TSV models with different ordering of the
returns lead to different posterior inference on the conditional
covariances. The differences in the dynamics of conditional
correlations are understandable because of the structure of the
conditional covariance matrix. In the TSV models the conditional
covariance between $\xi_{1,t}$ and $\xi_{2,t}$ (similarly between
$\xi_{1,t}$ and $\xi_{3,t}$) depends on the variance of
$\xi_{1,t}$ (i.e. $q_{11,t}$). Thus, a large increase in the
conditional variance of $\xi_{1,t}$ leads to an increase in the
conditional covariance. Therefore the $TSV_{WSF}$ and $TSV_{WFS}$
models (in which the WIG index is the first component) lead to
similar inference on the dynamics of the conditional correlations.
The plots of the posterior means of $\rho_{ij,t}$, obtained in the
remaining TSV models are different (because of differences in
volatilities of the S\&P$500$, FTSE indexes and WIG index). Note
also that in the JSV model the latent processes that describe
volatilities are included in the conditional correlation
coefficient definitions. Consequently, the conditional
correlations depend on the volatilities. Surprisingly, in the SDF
model the conditional correlations are estimated very precisely -
the posterior standard deviations of $\rho_{ij,t}$ are relatively
small. The returns on the WIG index are lower correlated with
returns on the S\&P $500$ index (with an average of $0.18$) than
with returns on the FTSE index (with an average of $0.24$). This
low correlation is partially explained by the non - overlapping
trading hours of U.S. market with the European markets. The U.S.
market (represented by the S\&P $500$ index) has the average
correlation of $0.47$ with the U.K. market. Finally, it is
important to stress that our results show that the conditional
correlations are not significantly higher when world markets are
down trending, which is in contrast to the results presented in
the papers: \cite{Ang}, \cite{Solnik}, \cite{Longin}.

\begin{figure}[t]\label{Figure:1}
  \includegraphics[width=13cm]{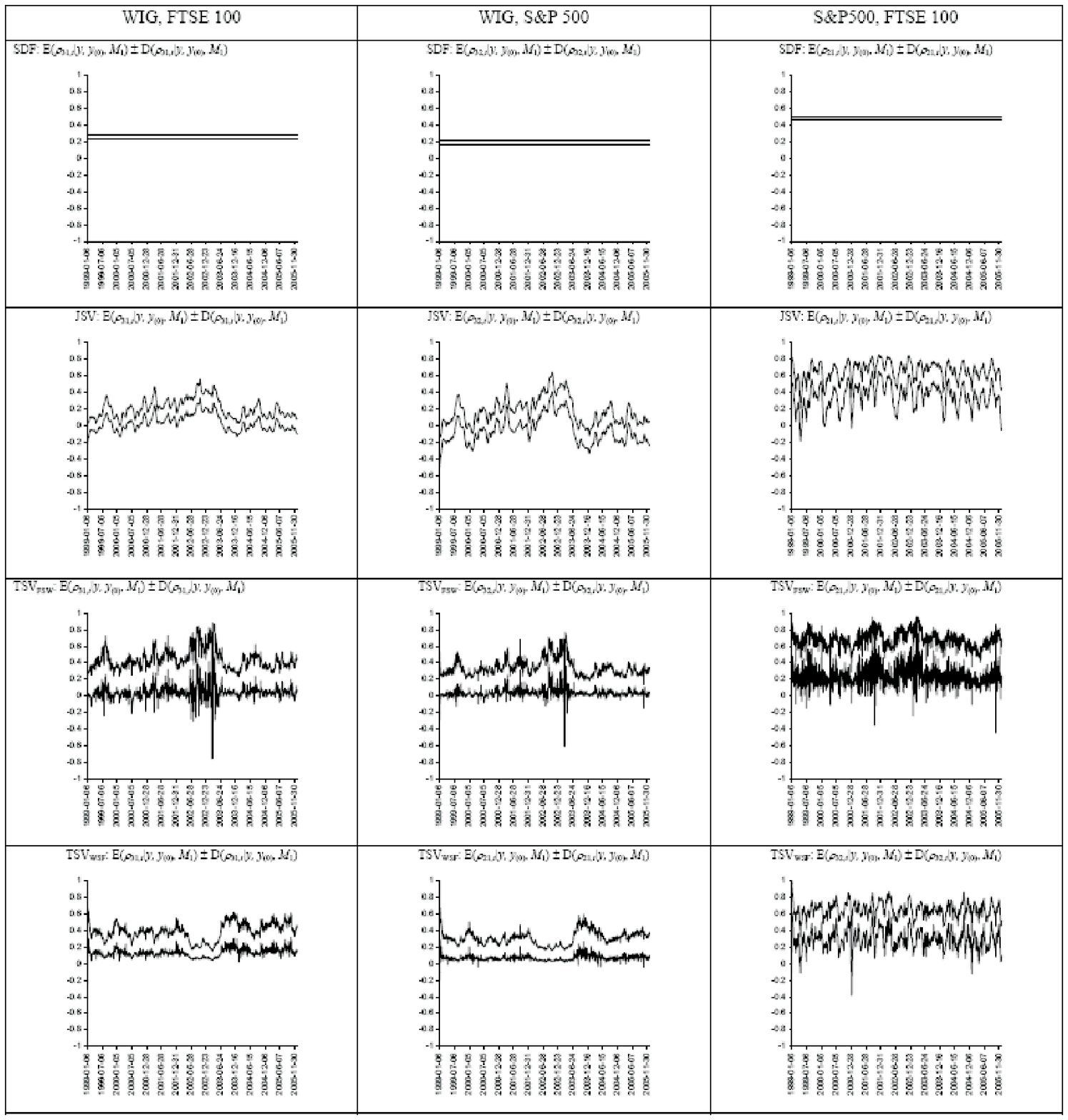}\caption{Conditional correlation coefficients (posterior mean $\pm
   1$ standard deviation)}
\end{figure}

\listoffigures
\listoftables


\begin{thebibliography}{99}
\bibitem{Ang}{Ang A., Bekaert G. (2002), International Asset Allocation
With Regime Shifts, The Review of Financial Studies 15, 1137-1187}

\bibitem{Harvey}{Harvey A. C., Ruiz E., Shephard N.G. (1994), Multivariate
Stochastic Variance Model, Review of Economic Studies, vol.61}

\bibitem{Gamerman}{Gamerman D. (1997), Markov Chain Monte Carlo. Stochastic
Simulation for Bayesian Inference, Champan and Hall, London}

\bibitem{Jacquier}{Jacquier E., Polson N., Rossi P., (1995), Model and Prior for
Multivariate Stochastic Volatility Models, technical report,
University of Chicago, Graduate School of Business}

\bibitem{Longin}{Longin F., Solnik B., (2001), Extreme Correlation of
International Equity Markets, The Journal of Finance, vol. 56, no.
2, 649-676}

\bibitem{Newton}{Newton M.A., Raftery A.E., (1994), Approximate Bayesian inference
by the weighted likelihood bootstrap (with discussion), Journal of
the Royal Statistical Society B, vol. 56, No. 1}

\bibitem{Hagan}{O'Hagan A. (1994) Bayesian Inference, Edward Arnold, London}

\bibitem{Osiewalski}{Osiewalski J., Pajor A., Pipieñ M. (2006) Bayes factors for
bivariate GARCH and SV models, Acta Universitatis Lodziensis - Folia
Oeconomica, forthcoming}

\bibitem{Pajor:2003}{Pajor A., (2003), Procesy zmiennoœci stochastycznej w
bayesowskiej analizie finansowych szeregów czasowych (Stochastic
Volatility Processes in Bayesian Analysis of Financial Time Series),
doctoral dissertation (in Polish), published by Cracow University of
Economics, Kraków}

\bibitem{Pajor:2005a}{Pajor A., (2005a), Bayesian Analysis of Stochastic Volatility
Model and Portfolio Allocation, [in:] Issues in Modelling,
Forecasting and Decision-Making in Financial Markets, Acta
Universitatis Lodzensis - Folia Oeconomica 192, 229-249}

\bibitem{Pajor:2006}{Pajor A. (2006), VECM-TSV Models for Exchange Rates of the Polish
Zloty, [in:] Issues in Modelling, Forecasting and Decision-Making in
Financial Markets, Acta Universitatis Lodzensis - Folia Oeconomica,
forthcoming}

\bibitem{Solnik}{Solnik B, Boucrelle C., Fur L. Y., (1996), International Market
Correlation and Volatility, Financial Analysis Journal vol.52, no.5,
17-34}

\bibitem{Tsay}{Tsay R.S., (2002), Analysis of Financial Time Series. Financial
Econometrics, A Wiley-Interscience Publication, John Wiley \& Sons,
INC}

\end{thebibliography}
\end{document}